# Suppressing transverse mode instability through multimode excitation in a fiber amplifier

Chun-Wei Chen,[1,3] Kabish Wisal,[2,3] Yaniv Eliezer,[1] A. Douglas Stone,[1] Hui Cao[1,*]

[1]Department of Applied Physics, Yale University, New Haven, CT 06520, USA
[2]Department of Physics, Yale University, New Haven, CT 06520, USA
[3]Contributed equally to this work
[*]Corresponding author: hui.cao@yale.edu

## Abstract

High-power fiber laser amplifiers have enabled an increasing range of applications in industry, medicine and defense. The power scaling for narrow-band amplifiers is currently limited by the transverse modal instability. Various techniques have been developed to suppress the instability in a single or few-mode fiber in order to output a clean, collimated beam. Here we propose to use a highly multimode fiber and equal modal excitation to suppress the thermo-optical nonlinearity and instability. Our numerical simulations and theoretical analysis predict a significant reduction of dynamic coupling among the fiber modes with such excitation. When the bandwidth of a coherent seed is narrower than the spectral correlation width of the multimode fiber, the amplified light maintains high spatial coherence and can be transformed to any target pattern or focused to a diffraction-limited spot by a spatial mask at either input or output end of the amplifier. Our method simultaneously achieves high average power, narrow spectral width, and good beam quality, which are desired for fiber amplifiers in many applications.

# Introduction

The ever-increasing demands for high-power lasers in current and emerging technologies have driven rapid advances in fiber laser amplifiers, leading to an enormous progress in the past twenty years[1-7]. The master-oscillator-power-amplifier (MOPA) architecture combines a seed laser with several stages of fiber amplifiers to output more than a kilowatt of power continuously[8]. While the double-clad fiber structure allows cladding-pumping with high-power semiconductor laser diodes, the single-mode core ensures good beam quality of the amplified light[9]. However, the high intensity of light confined in a small core plus the long propagation path in the fiber enhances nonlinear effects that degrade the laser performance[10].

For continuous-wave fiber amplifiers, the maximum power is currently limited by the transverse mode instability (TMI)[2,3,11]. It originates predominantly from the thermo-optical nonlinearity, in the following manner[12-28]. Even in a single-mode fiber amplifier, thermal or nonlinear effects can create and excite additional fiber modes. The interference between the fundamental mode (FM) and a higher-order mode (HOM) generates an optical intensity grating. This results in nonuniform quantum-defect heating, producing a refractive-index grating through the thermo-optical effect. The dynamic part of the index grating results in power transfer between fiber modes. This dynamic mode coupling causes a sudden onset of fluctuations in the output beam profile above a power threshold, which significantly reduces the spatial coherence and the beam quality[29,30].

Over the past decade, many efforts have been devoted to mitigate TMI. Minimizing the HOM amplification has become one of the main design goals in such fiber amplifiers, to improve their thermo-optical stability[31-33]. An alternative approach to suppressing TMI is to reduce the FM–HOM coupling by smearing out the thermally-induced dynamic refractive-index grating[34-36]. Two methods, adaptive control over the seed, or pump modulation rely on a feedback system to stabilize the output beam profile[34,37]. Additional methods of mitigating the TMI include: synthesizing new fiber materials with low thermo-optical coefficient, shifting pump or seed wavelength, modifying the pump configuration, coiling the fiber, and dividing the power in an array of single-mode cores[38-48]. Almost all the approaches strive to maintain stable single-mode amplification for good output beam quality.

Here we introduce an efficient, robust mechanism for TMI suppression based on coherent many-mode amplification in a multimode fiber. Our numerical simulations and theoretical modeling demonstrate a dramatic increase of the TMI threshold with equal excitation of many HOMs. Even in a two-mode fiber amplifier, equally distributing the seed power between the two modes greatly suppresses the dynamic mode coupling, as the nonlinear thermo-optical gain for one mode depends strongly on the power in the other mode. By an extension of this reasoning, we expect (and will show) that the TMI suppression is further enhanced by coherent excitation of additional HOMs. Physically, the modal interference generates a complex spatial modulation of light intensity with fine speckles. Hence the local heating varies on this scale, but slow thermal diffusion results in a much smoother temperature distribution, which varies on a large length scale. This spatial mismatch between temperature and optical intensity variation drastically reduces the number of fiber modes with strong thermo-optical coupling to those with small difference between their propagation constants. Consequently, we show that the TMI threshold increases linearly with the number of equally-excited modes in a fiber amplifier.

Maintaining the output beam quality is a major concern for such multimode fiber amplifiers.

Our dynamical simulations shows that below the TMI threshold, as long as the seed linewidth is narrower than the fiber spectral correlation width, the amplified light in individual modes remains mutually coherent. Then a single spatial mask placed at the fiber output end can be used to convert highly speckled light to a collimated Gaussian beam. Alternatively, the mask may be placed at the fiber input end to shape the seed wavefront for focusing the output beam to a diffraction-limited spot[49], even in the presence of thermo-optical nonlinearity. Therefore, our method allows stable, ultrahigh power generation in a many-mode fiber amplifier while maintaining high spatial and temporal coherence. This multimode approach may also be applied to suppress other nonlinear optical effects in fiber amplifiers, such as stimulated Brillouin and Raman scattering[50].

# Results

## Numerical simulation

Due to computational constraints, we simulate the thermo-optically-induced TMI in a two-dimensional (2D) multimode waveguide with linear optical gain. The waveguide has a core of width $W_c = 40$ μm, and refractive index $n_s = 1.5$. Its cladding is 400 μm wide, and its length is $L = 1$ m. Monochromatic radiation at wavelength $\lambda_s = 1064$ nm is launched into the waveguide as the seed. The pump wavelength is $\lambda_p = 975$ nm. Ignoring the evanescent wave in the cladding, the $m$-th guided mode in the core has the transverse (along $x$ axis) field profile $\phi_m(x) = \sin[m\pi(x + W_c/2)/W_c]$, and the longitudinal (along $z$ axis) propagation constant $\beta_m = \sqrt{k^2 - (m\pi/W_c)^2}$, where $m = 1, 2, ...$, and $k = 2\pi n_c/\lambda_s$. The total field in the waveguide is decomposed as $\psi(x, z, t) = \sum_m A_m(z, t)\phi_m(x)$, where $t$ denotes time. The scalar paraxial optical wave equation gives

$$\frac{\partial}{\partial z} A_m(z, t) = \left(i\beta_m + \frac{g}{2}\right) A_m(z, t) + ik \sum_{j \neq m} \left[A_j e^{i(\beta_j - \beta_m)z} \int_{-W_c/2}^{+W_c/2} \phi_j^*(x) \Delta n_c(x, z, t) \phi_m(x)\, dx\right], \quad (1)$$

where $g$ is the mode-independent linear-gain coefficient, and $\Delta n_c(x, z, t)$ is the thermally-induced refractive-index change in the core. The last term on the right-hand side of Eq. (1) represents nonlinear mode coupling, while linear mode coupling and optical gain saturation are neglected.

The refractive-index change induced by the non-uniform heating is given by $\Delta n_c(x, z, t) = (dn/dT)\Delta T(x, z, t)$, where $dn/dT$ is the thermo-optical coefficient and $\Delta T = T(x, z, t) - T(x, z, 0)$ is the local deviation of the temperature from its value at $t = 0$. The latter is obtained by solving the heat diffusion equation:

$$\rho C \frac{\partial}{\partial t} T(x, z, t) = Q(x, z, t) + \kappa \left(\frac{\partial^2}{\partial x^2} + \frac{\partial^2}{\partial z^2}\right) T(x, z, t), \quad (2)$$

where $\rho$, $C$, $\kappa$ are the mass density, specific heat capacity, thermal conductivity of the waveguide, respectively. The quantum-defect heating $Q(x, z, t) = g(\lambda_s/\lambda_p - 1)I(x, z, t)$, where $I(x, z, t) = n_c|\sum_m A_m(z, t)\phi_m(x)|^2$ is the local intensity. The outer boundaries of the cladding are set to be perfectly thermally conducting (Dirichlet boundary condition).

With given input field amplitude and phase in each waveguide mode, the coupled optical and thermal equations are solved numerically in the time domain (see Methods for details). In practice, there is always broadband noise, which commonly arises from the power fluctuation of the seed laser. After finding the steady-state solution for a time-invariant input signal, we add a perturbation—temporal fluctuation at frequency $\Omega$—to the seed to simulate the effect of the noise. We assume that the noise power in each mode is proportional to the time-invariant signal power. We compute the temporal fluctuation of the output field and examine its dependence on the perturbation frequency $\Omega$.

## Two-mode amplifier

We first show how dynamic mode coupling can be suppressed by equal modal excitation. For simplicity, we consider only two modes, the FM and a HOM, excited with distinct power ratios at the input: (I) FM:HOM = 99:1, (II) FM:HOM = 50:50. (I) corresponds to the typical situation that a HOM is created in a single-mode fiber at high power due to thermal and/or nonlinear effects, and excited slightly either at the fiber input or via coupling with the FM. The effective number of excited modes is $M_e \approx 1$ in (I) and $M_e = 2$ in (II). Ignoring linear mode coupling and mode-dependent gain, the signal mode content is retained throughout the waveguide. For the FM-dominant excitation (I), the steady-state light intensity $I_{ss}(x,z)$ and temperature $T_{ss}(x,z)$ distributions, in Fig. 1a and 1c, both exhibit a small undulation on top of the FM background, as a result of weak modal interference. In the case of equal-mode excitation (II), a 50-times increase in the HOM content leads to a much stronger contrast of light intensity grating in Fig. 1b. The temperature grating in Fig. 1d is of the same period and shows a similarly high contrast. However, the resulting refractive-index grating is static in time and does not cause any dynamic power transfer between the modes.

Next a sinusoidal perturbation at $\Omega = 1$ kHz is added to the input field, with strength $3 \times 10^{-5}$ of the signal power. With the FM-dominant excitation, both intensity and temperature distributions start changing in time [see Fig. 1e and Supplementary movie 1], leading to a moving refractive-index grating, which enables dynamic power transfer between the modes. For the FM-dominant excitation, even though the initial HOM content is very low, it grows rapidly along the waveguide due to nonlinear coupling with the FM [Fig. 1g]. Close to the waveguide end, the FM and HOM contents exhibit strong anti-phase oscillations in time, leading to strong fluctuations of the output beam profile in Fig. 1i. Such dynamic power transfer between the FM and HOM is clearly indicative of TMI. In sharp contrast, when the two modes are equally excited with the same total power at the waveguide input, the dynamic part of the temperature grating is much weaker [Fig. 1f vs. 1e, note change of color scale], suppressing dynamic mode coupling and power transfer as shown in Fig. 1h. Consequently, at the same power level, the output beam profile remains stable [Fig. 1j]. Therefore the thermo-optical stability of the amplifier is greatly improved by evenly distributing the seed power between the modes.

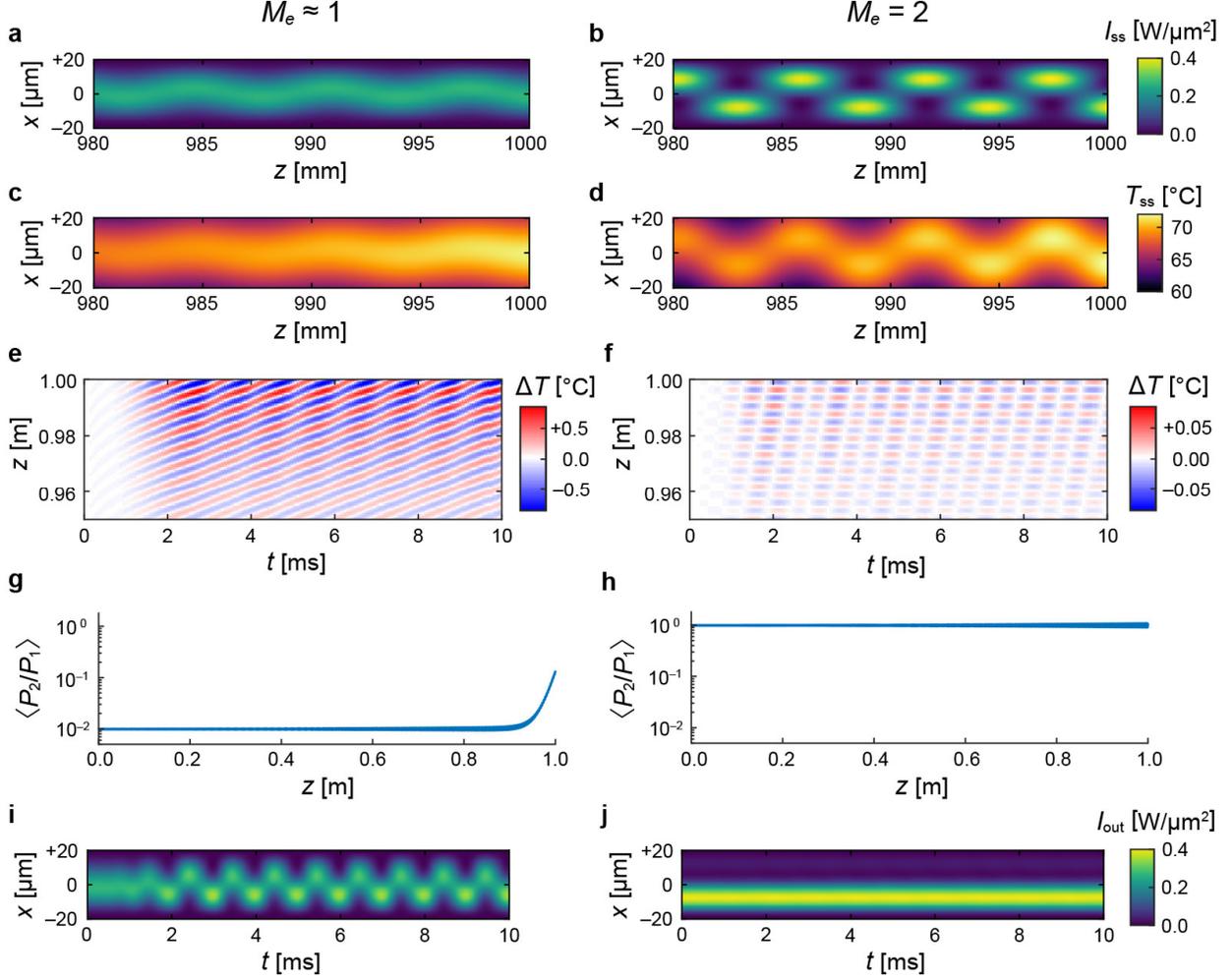

**Fig. 1 Two-mode amplification.** The input power ratio of the FM and HOM is 99:1 in the left column ($M_e \approx 1$), and 50:50 in the right column ($M_e = 2$). Panels **a–d** have no perturbation (noise) added to the input signal, **e–j** have a sinusoidal perturbation at 1 kHz. The standard deviation of temporal fluctuations of the input power in each mode is $3 \times 10^{-5}$ of the time-invariant signal power. **a,b** Steady-state light intensity distribution $I_{ss}(x,z)$ with $M_e \approx 1$ in **a** displays much weaker mode interference than that of $M_e = 2$ in **b**. The input and output powers are 23 W and 206 W, respectively. **c,d** Steady-state temperature distribution $T_{ss}(x,z)$ with $M_e \approx 1$ in **c** has much lower contrast than that of $M_e = 2$ in **d**. **e,f** The temperature change $\Delta T = T(z,t) - T(z,0)$ (at $x = 10$ μm, $z$ near the output) induced by the input perturbation is much larger for $M_e \approx 1$ in **e** than $M_e = 2$ in **f**. Note that the scale for $\Delta T$ in **e** is increased 10 times compared to **f**. **g,h** Power ratio of HOM and FM (averaged over $t = 5$–10 ms) increases rapidly along the waveguide with $M_e \approx 1$ in **g**, while it stays nearly invariant throughout the waveguide for $M_e = 2$ in **h**. **i,j** Output beam profile with $M_e \approx 1$ begins to oscillate after 1 ms of perturbation in **i**, indicative of the onset of TMI. In contrast, the output profile with $M_e = 2$ remains almost unchanged over time in **j**, and its shape is determined by the relative phase of input fields in the two modes.

To gain physical understanding of the time-domain simulation results, we resort to a semi-analytic theory of TMI in the frequency domain[17,19]. As in the simulation, we consider a monochromatic (frequency $\omega_0$) signal of fixed total power that is launched into the FM and a HOM in a certain ratio, along with noise at frequency $\Omega$ proportional to the signal power in each mode. The interference between the signal and noise creates a moving intensity grating. Through quantum defect heating and the thermo-optical effect, a moving refractive-index grating is generated and induces power transfer between the signal and noise. Two sidebands are formed at $\omega_0 \pm \Omega$. The downshifted component ($\omega_0 - \Omega$) gains power from signal and manifests as output-power fluctuation, while the upshifted component ($\omega_0 + \Omega$) is suppressed. We thus focus on the fluctuation at $\omega_0 - \Omega$ in the HOM, which grows exponentially, owing to both linear optical gain and thermo-optical gain, according to[19]:

$$\tilde{P}_2(z) = \tilde{P}_2(0) \cdot e^{gz} \cdot e^{\chi_{21}(\Omega)[P_1(z) - P_1(0)]}, \tag{3}$$

where $\tilde{P}_2(z)$ denotes the noise power in the HOM at $\omega_0 - \Omega$, $P_1(z)$ is the signal power in the FM, and $\chi_{21}(\Omega)$ is the nonlinear susceptibility due to thermo-optically induced coupling between the FM and HOM at frequency downshift $\Omega$ (see Methods). The final factor on the right-hand side of Eq. (3) shows that the exponential growth rate of the noise power increases *linearly* with the signal power in the FM. This expression is valid below the TMI threshold (ignoring non-phase-matched terms and the interaction of noise and signal in the same mode, which occurs at much lower frequencies[17]).

A critical insight from Eq. (3) is that the HOM power fluctuation depends linearly on its own input noise power $\tilde{P}_2(0)$, which is proportional to its signal power $P_2(0)$, but it also depends *exponentially* on the thermo-optical gain, which is proportional to the FM signal power $P_1(z)$. Hence the noise power at the output, $z = L$, is much more sensitive to the latter. Therefore, trading less input noise in the HOM for its slower growth rate is highly favorable for reducing TMI. Specifically, for the parameters relevant to our simulation, changing from the FM-dominant to equal-mode excitation increases the HOM input noise power $\tilde{P}_2(0)$ by 50 times., but the halving of the FM signal power $P_1(0)$ in the exponent leads to a reduction of the final term in Eq. (3) by $3.7 \times 10^5$. Consequently, the HOM noise power at the waveguide end, $\tilde{P}_2(L)$, is $1.35 \times 10^{-4}$ smaller with equal-mode excitation than with FM-dominant excitation.

## Multimode excitation

The previous section shows that equal power division between two modes effectively suppresses TMI. The natural question arises: will TMI be further suppressed if even more modes are excited? The answer is not obvious a priori. When the seed power is divided among multiple modes, thermo-optical coupling between any mode pair is weaker, but there are more mode pairs. The increase in the number of coupled mode pairs might counter the decrease of coupling between each mode pair. The noise in one mode may acquire thermo-optical gain from signals in all other modes, despite the gain from each individual mode being lower. However, we will show below that only a fraction of total mode pairs have strong thermo-optical coupling, leading to significant suppression of the TMI upon many-mode excitation.

Below we consider five-mode amplification in the 2D waveguide. A monochromatic seed coherently excites the five lowest-order modes equally. The multimodal interference produces a fine-grained intensity pattern $I_{ss}$, as shown in Fig. 2a. The first and fifth modes have the largest difference in transverse and longitudinal wave numbers, and their beating periods are the shortest in both directions, e.g., their longitudinal beating period is 7 times shorter than that between the first and second modes. However, the fine grains in the light intensity pattern $I_{ss}$ disappear in the steady-state temperature distribution $T_{ss}$ in Fig. 2b. This is because the heat diffusion occurs on a much longer scale and smears out the short-scale intensity variations that act as local heat sources.

This effect is evident when comparing the longitudinal Fourier spectra of $I_{ss}$ and $T_{ss}$ in Fig. 2c. The high-frequency peaks in $I_{ss}$ due to longitudinal beating of non-adjacent modes are strongly suppressed in $T_{ss}$, which contains mostly the low-frequency peaks from adjacent-mode beating. The resulting refractive-index grating couples mostly adjacent modes, and the thermo-optical interaction between non-adjacent modes is greatly damped.

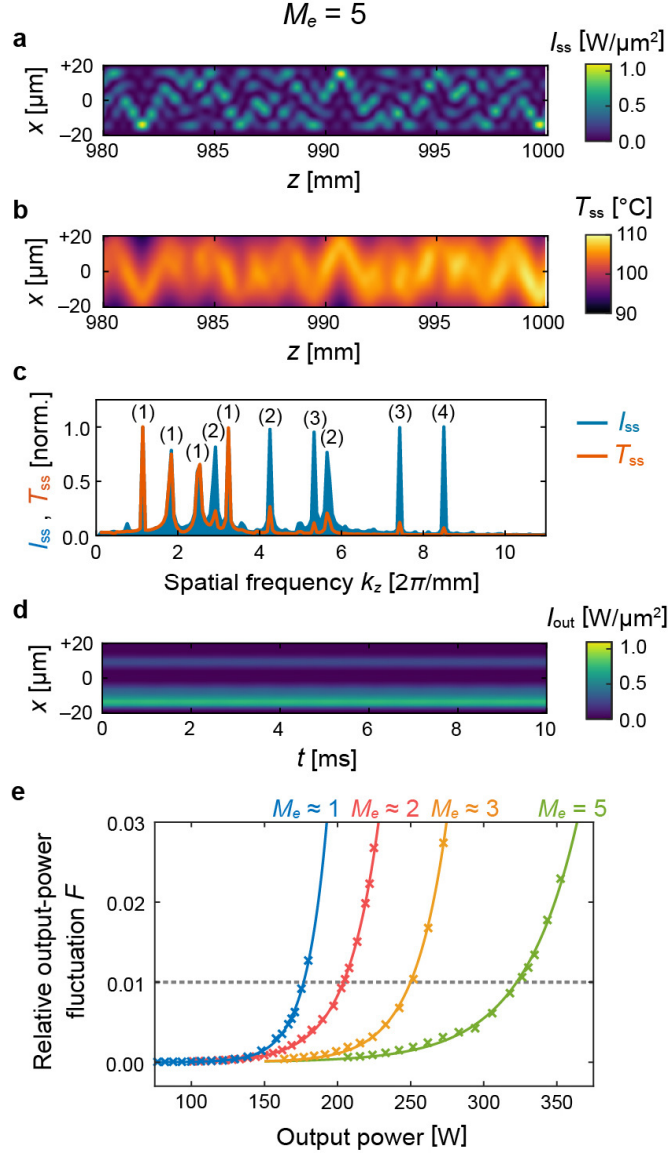

**Fig. 2 Five-mode amplification. a–d** $M_e = 5$, equal excitation of five lowest-order guided modes. **a** Steady-state light intensity distribution $I_{ss}(x, z)$ displays multimode interference. The input and output powers are 35 W and 315 W, respectively. **b** Steady-state temperature distribution $T_{ss}(x, z)$ exhibits much smoother and longer modulation than $I_{ss}(x, z)$. **c** Fourier spectrum of longitudinal intensity $\langle |I_{ss}(x, k_z)| \rangle_x$ (blue) consists of peaks from mode-beating. The black (#) above each peak denotes the spacing $|m - j|$ of corresponding mode pair. The high-frequency peaks from non-adjacent mode pairs are strongly suppressed in the longitudinal spectrum of temperature $\langle |T_{ss}(x, k_z)| \rangle_x$ (orange). **d** Output beam profile under input perturbation ($\Omega = 1$ kHz) remains stable in time. The standard deviation of temporal fluctuation of input power in each mode is $3 \times 10^{-5}$ of the signal power. **e** Relative output-power

fluctuations for different numbers of equally-excited modes $M_e$. The fluctuation $F$ grows exponentially with average output-power $\sum_m P_m(L)$ and with smaller growth rate for larger $M_e$, while input fluctuation is kept the same. Crosses are numerical data, and lines represent exponential fitting. Gray dotted line at $F = 0.01$ marks the TMI threshold, which increases with $M_e$.

Next we introduce an input perturbation at $\Omega = 1$ kHz to all five modes. While the ratio of the total input perturbation power to the total signal power is kept the same as that for two-mode excitation, the signal output-power is raised to 315 W by increasing the seed power. At this higher power level, as discussed below, TMI occurs under two-mode equal excitation, but the five-mode amplification remains stable, and the output pattern barely changes over time in Fig. 2d, indicating that the TMI has been further suppressed.

To further demonstrate the importance of equal excitation of all modes, we simulate different power ratios for the five modes: (i) 96:1:1:1:1; (ii) (97/2):(97/2):1:1:1; (iii) (98/3):(98/3):(98/3):1:1; (iv) 20:20:20:20:20. The effective numbers of excited modes are $M_e \approx 1,2,3$ and $M_e = 5$ in the four cases. The total input-power is kept constant, and the noise power is proportional to the signal power in each mode. For quantitative comparison of amplifier stability, we compute the relative output-power fluctuation:

$$F \equiv \frac{\sum_m \sigma_m}{\sum_m P_m(L)}, \tag{4}$$

where $\sigma_m$ is the standard deviation of temporal fluctuation of output-power in the $m$-th mode, and $\sum_m P_m(L)$ is the total output-power (in all modes) averaged over time $t$. Figure 2e reveals that $F$ grows exponentially with increasing $\sum_m P_m(L)$. The growth rate drops rapidly as $M_e$ increases from 1 to 5. We also vary the input perturbation at a fixed output-power, and show in Supplementary Information [Fig. S1] that the amplifier becomes much more robust against input perturbation with more-mode excitation.

To understand these results, we develop the frequency-domain theory of TMI for multimode excitation, by generalizing the approach of refs. 17,19. This theory implies that the noise power in the $m$-th mode grows exponentially as

$$\tilde{P}_m(z) = \tilde{P}_m(0) \cdot e^{gz} \cdot e^{\sum_{j \neq m} \chi_{mj}(\Omega)[P_j(z) - P_j(0)]}. \tag{5}$$

The thermo-optical susceptibility is given by:

$$\chi_{mj}(\Omega) = \alpha \sum_l \frac{\Omega}{\Omega^2 + \Gamma_l^2} \left| \int_{-W_c/2}^{+W_c/2} \psi_m^*(x) \psi_j(x) T_l(x) \, dx \right|^2, \tag{6}$$

where $\alpha$ is a material-dependent coefficient (see Methods), $T_l(x)$ is the transverse temperature profile for the $l$-th eigenmode of the heat diffusion equation, $\Gamma_l$ is its decay rate, which is on the order of a few kHz. For the five-mode waveguide, we calculate the nonlinear gain susceptibilities $\chi_{mj}$ of all mode pairs. For a given number of equally-excited modes $M_e$, the effective thermo-optical susceptibility for mode $m$ is defined as $\bar{\chi}_m \equiv \sum_{j \neq m} \chi_{mj}/M_e$, where $1/M_e$ is the fraction of signal power in each of the other modes. Finally, in Fig. 3a, we plot the average of $\bar{\chi}_m$ over all $m$, denoted by $\bar{\chi}$ (see Methods for details of the calculation). The value of $\bar{\chi}$ depends on the perturbation frequency $\Omega$, and reaches its maximum at around 1 kHz. Five-mode excitation ($M_e = 5$) greatly reduces the thermo-optical susceptibility, compared to FM-dominant excitation ($M_e \approx 1$). Figure 3b shows the maximal value of $\chi_{mj}$ for each pair of modes. The thermo-optically induced mode coupling is the strongest for the adjacent modes which have the smallest difference in propagation constant $\Delta\beta_{mj} = |\beta_m - \beta_j|$. $\chi_{mj}$ decreases monotonically with increasing mode spacing $|m - j|$. This short-ranged coupling is characteristic of a diffusion-mediated process; physically it arises, as noted above, from the mismatch in length scale between thermal diffusion and optical interference. This enters the theory through the overlap integral in Eq. (6), which dictates that only the thermal mode $l$ with spatial scale matching the beating period of two optical modes $m$ and $j$ can cause significant coupling between $m$ and $j$. The contribution to $\chi_{mj}$ from thermal mode $l$ peaks at frequency downshift $\Omega = \Gamma_l$, and the peak value is proportional to the mode lifetime $1/\Gamma_l$. Thus for a pair of optical modes with large spacing $|m - j|$, their beating period $1/\Delta\beta_{mj}$ is short, so only the thermal mode with a high spatial frequency (large $l$) contributes to the thermo-optical coupling. However, such a thermal mode has a short lifetime, $1/\Gamma_l$, leading to weak coupling strength for optical modes with substantially different propagation constants.

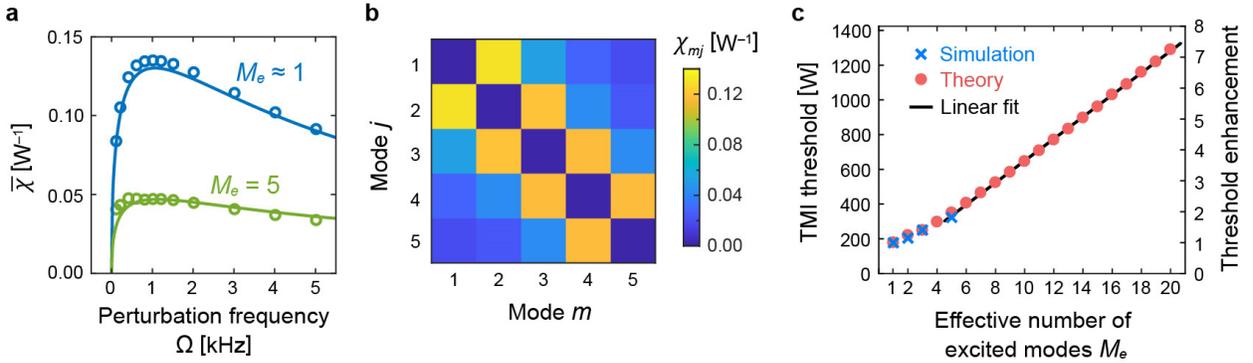

**Fig. 3 Thermo-optical mode coupling and TMI threshold. a** Effective thermo-optical susceptibility $\overline{\chi}$ for $M_e = 5$ and $M_e \approx 1$ varies with frequency downshift $\Omega$ from sinusoidal perturbation and peaks around 1 kHz. The maximum of $\overline{\chi}$ for equal excitation of five modes ($M_e = 5$) is $\sim M_e/2 = 2.5$ times weaker than that under FM-dominant excitation ($M_e \approx 1$). Circles are numerical data from time-domain simulation, lines from semi-analytic frequency-domain theory. **b** Maximal value of $\chi_{mj}$, for every pair of modes ($m \neq j$), decreases monotonically with increasing mode spacing $|m - j|$. **c** TMI threshold increases linearly with effective number of excited modes $M_e$. Blue crosses are time-domain simulation results, red circles are predictions of frequency-domain theory, black line is a linear fit with slope 0.36 for the threshold enhancement factor over the FM-dominant excitation ($M_e \approx 1$) for $5 \leq M_e \leq 20$.

This short-ranged nature of the thermo-optical coupling ensures that the noise growth rate in any waveguide mode is effectively determined by the signal power in its two neighboring modes. Therefore, if the signal power is equally divided into $M_e \gg 2$ modes, the noise growth rate in Eq. (5) drops by a factor of $\sim M_e/2$. Alternatively, for the same noise at the output, the signal power is raised by a factor of $\sim M_e/2$. Note that non-adjacent mode couplings are weak but non-zero, thus the slope of enhancement factor is slightly less than 1/2.

In Fig. 3c, we evaluate the TMI threshold quantitatively for the 2D waveguides which we have simulated. Defining the threshold as the signal output power when the relative output-power fluctuation reaches $F = 0.01$, we find that threshold output-power does scales linearly with $M_e$ for $M_e \geq 5$, with slope $\approx 0.36$. The exact simulation results agree with the theory quantitatively, with no fitting parameters, for $M_e \leq 5$; for $M_e > 5$ the simulations are impractical but the theory (which requires only numerical evaluation of the overlap integrals in Eq. (6)) is straightforwardly extended to any relevant number of modes. We define a threshold enhancement factor as the ratio of the threshold for $M_e$ equally excited modes to that under the FM-dominant excitation ($M_e \approx 1$). We find that choosing $M_e = 20$ leads to a seven-fold increase of the TMI threshold [Fig. 3c].

## Output beam quality

One serious negative consequence of dynamical mode coupling is the resulting temporal fluctuations of the transverse field profile, which corresponds to a loss of spatial coherence. While multimode excitation itself produces a speckled field, the spatial coherence is not necessarily lost. Let us consider a multimode waveguide whose output field pattern $E_{\text{out}}(x, \lambda)$ varies with the input wavelength $\lambda$. The spectral correlation function is defined as $C(\Delta\lambda) \equiv \langle|\langle E^*_{\text{out}}(x,\lambda)E_{\text{out}}(x,\lambda+\Delta\lambda)\rangle_x|\rangle_\lambda$, where $\langle...\rangle_x$ denotes averaging over $x$, and $\langle...\rangle_\lambda$ over $\lambda$. $C(\Delta\lambda)$ decays with $\Delta\lambda$, and its width at half maximum is defined as the spectral correlation width.

If the input light has a bandwidth narrower than the spectral correlation width of a multimode waveguide, the output field pattern stays fairly constant in time, except for a global amplitude and/or phase drift. Thus the output field is spatially coherent, even if it displays a complex interference pattern. This will be reflected in the degree of spatial coherence: $C_s \equiv \langle|\langle E^*_{\text{out}}(x,t)E_{\text{out}}(x,t+\Delta t)\rangle_x|\rangle_{t,\Delta t}$, where the output field $E_{\text{out}}(x,t)$ is normalized so that $\int_{-W_c/2}^{+W_c/2}|E_{\text{out}}(x,t)|^2\,\mathrm{d}x = 1$ at any $t$.

Figure 4 shows the degree of spatial coherence of the output field for different modal excitation in a five-mode amplifier. Below the TMI threshold, $C_s \simeq 1$ reflects nearly-perfect spatial coherence in the presence of small perturbation. Above the TMI threshold, $C_s$ drops quickly, as the output pattern changes with time.

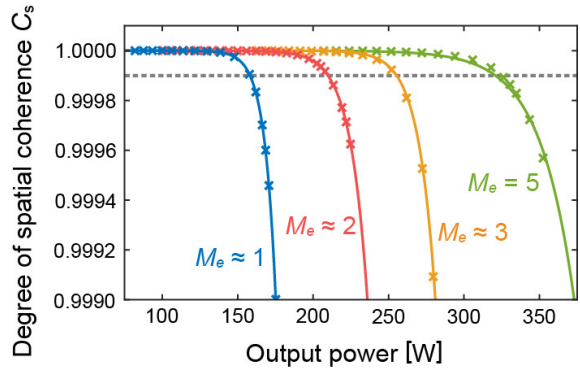

**Fig. 4 Spatial coherence of multimode amplifier.** Degree of spatial coherence of the output field $C_s$ vs. average output-power $\sum_m P_m(L)$ reflects near-perfect spatial coherence below the TMI threshold (marked by gray dotted line) but degrades quickly above the threshold. The standard deviation of temporal fluctuations of the input power in each mode is $3 \times 10^{-5}$ of the signal power. At a fixed average output-power, $C_s$ is higher for larger effective number of excited modes $M_e$.

Below the TMI threshold, since the output field pattern $E_{\text{out}}(x)$ is constant in time, a single spatial mask placed at the amplifier output end can convert it to any desired pattern $E_d(x) = T(x) E_{\text{out}}(x)$, where $T(x)$ is the field transmission coefficient of the spatial mask. In practice, the high output power of an amplifier may damage the mask. Alternatively, the mask may be placed at the amplifier input end to shape the spatial wavefront of a coherent seed in order to create a desired output profile.

Since our approach to suppressing TMI relies on equal excitation of all waveguide modes, it does not impose any restriction on the phase of the input field in each individual mode. Therefore, the input phases of the excited modes may be adjusted to shape the output beam profile. As an example, we demonstrate, in the numerical simulation, focusing the output beam to a diffraction-limited spot by wavefront shaping of the input light, while simultaneously achieving a high TMI threshold. In Fig. 5, we show a 270-W output beam from the five-mode amplifier focused to a diffraction-limited spot, using the input field amplitude and phase in each mode shown in Fig. 5a. A sharper and cleaner focus than Fig. 5c should be attainable in a fiber with more modes available. Figure 5d reveals that the output intensity pattern is invariant over time, with the spatial coherence $C_s = 0.9999$. This performance contrasts with that of the conventional quasi-single-mode amplifier; its output fluctuates drastically in time at the same power level (not shown). Compared to the 153 W threshold for the FM-dominant excitation, the TMI threshold in the case of output focusing with five modes reaches 270 W.

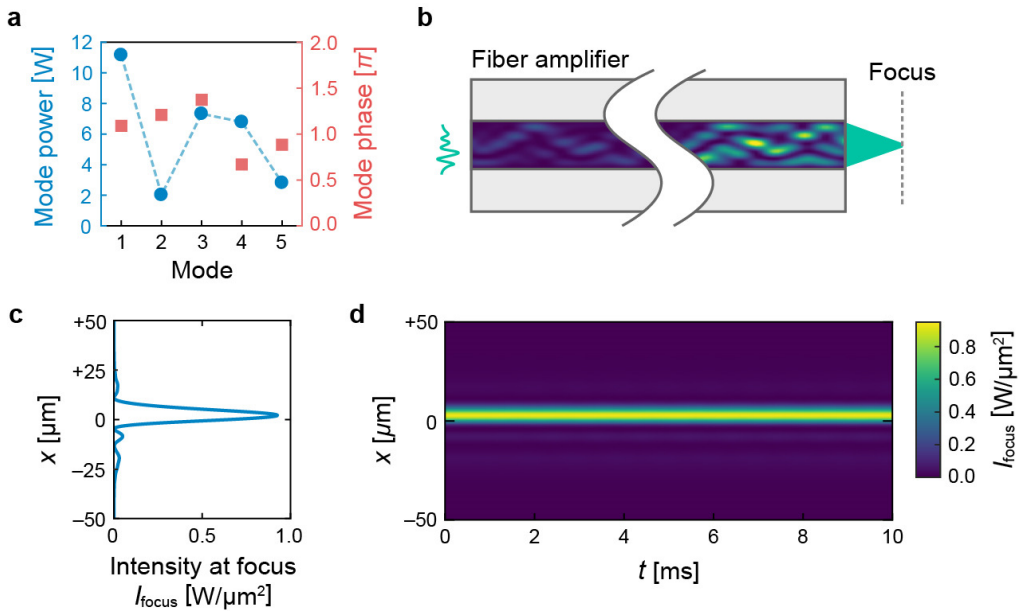

**Fig. 5 Adjusting seed wavefront to focus through multimode amplifier. a** Power and phase of input light in each of five modes in the waveguide. **b** Schematic of focusing the output light to a spot at 174 μm away from the end facet and 2.2 μm off the waveguide axis. **c** Transverse intensity distribution at focal plane has a diffraction-limited width of 10.4 μm, given by the full width at $e^{-2}$ of the maximum intensity. **d** Output beam profile at the focal plane remains stable in time under input perturbation ($\Omega = 1$ kHz). The standard deviation of temporal fluctuations of the input power in each mode is $3 \times 10^{-5}$ of the signal power.

## Discussion

In summary, equal excitation of many guided modes stabilizes a fiber amplifier against the thermo-optical nonlinearity. The maximum growth rate for the noise power leading to TMI is reduced by distributing the power into many modes, because any given mode only experiences strong gain from signal power in the adjacent modes (i.e., modes with the closest propagation constants). "Non-adjacent modes" have weak thermo-optical coupling and contribute little to the dynamic power transfer. The substantially reduced coupling strength of non-adjacent modes is due to the large mismatch of thermal and optical length scales. The linear scaling of the TMI threshold with the number of equally excited modes provides a powerful path toward robust TMI suppression, as a multimode fiber (MMF) can easily support hundreds of modes. In principle, the threshold enhancement is only limited by the total number of modes in a fiber.

We also demonstrated that the output beam profile of a MMF amplifier with thermo-optical nonlinearity can be tailored by wavefront shaping of a coherent seed with a spatial mask. The ability of focusing, collimation, and shaping the output beam is crucial for applications such as laser machining and processing. Several free-space or all-fiber techniques[37,49] are readily available for experimental realization of shaping a CW seed up to ~100 W.

Although the current numerical and theoretical study is conducted on a 2D waveguide with a 1D cross-section, our theoretical method is applicable to a standard fiber with a 2D cross-section. If polarization mixing is strong in a fiber, wavefront shaping must be applied to both polarizations of the input light for a full control of output beam profile. In our study, both linear mode coupling and mode-dependent gain/loss are neglected. While the latter reduces the effective number of excited modes, the former tends to spread power among all fiber modes and facilitate our equal-excitation scheme. Therefore, our TMI suppression method works well in the presence of fiber bending and imperfections, which induce linear mode coupling[46]. Another effect that is not accounted for in this study is optical gain saturation, which reduces heating and raises the TMI threshold[39,41]. We expect the threshold increase we have found under multimode excitation to be present even if gain saturation is taken into account. As noted above, the advantage of multimode excitation stems from the mismatch in the length scale for optical mode interference and the much larger scale over which the induced temperature variations occur. Because the characteristic scale of optical mode interference is not modified by gain saturation, it should not alter the short-range nature of thermo-optical coupling, which leads to the linear scaling of the TMI threshold with the effective number of excited modes.

Finally, our method may be combined with other schemes for TMI suppression. Since the mismatch between optical and thermal length scales is almost universal among MMFs, our multimode excitation scheme will work for specialty fibers that have been designed to increase the TMI threshold[32,33,47]. This also opens a vast space for fiber structure design to enable output-beam engineering from tailored mode structures or to offer new functionalities. We envision that, with optimized modal excitation, multimode fiber amplifiers can operate stably with high beam-quality at extreme powers that would be unachievable with their single-mode counterparts.

## Methods

### Time-domain numerical simulation

We choose the parameters in our simulation to be comparable with other numerical studies[14,17-19,25,27]. The thermo-optical coefficient is $dn/dT = 1.17 \times 10^{-5}$ °C$^{-1}$, the product of mass density and specific heat capacity is $\rho C = 1.67 \times 10^{-12}$ J K$^{-1}$ μm$^{-3}$, and the thermal conductivity is $\kappa = 1.4 \times 10^{-6}$ W K$^{-1}$ μm$^{-1}$. We ignore the pump depletion and assume the optical gain is constant for all modes throughout the waveguide. The power gain coefficient is $g \approx 2.2$ m$^{-1}$.

Since heat diffusion is much slower than light propagation through the amplifier, the temperature distribution throughout the waveguide is treated as static during the time that light travels the entire length of the amplifier[14]. At each time step, Eq. (1) is solved with a static $\Delta n_c$, which is given by the current temperature distribution. The resulting modal amplitudes along the waveguide are then used to calculate the heat load to update the temperature distribution. More specifically, we use the Crank–Nicolson scheme to solve Eq. (2) for the temperature distribution at the next time step. The longitudinal (along $z$) heat diffusion is neglected in our simulations, because the temperature gradient in $z$ is much smaller than that in $x$. Note that the longitudinal heat diffusion will become comparable to the transverse heat diffusion (along $x$), when a large number of modes are excited.

We carefully adjust the temporal and spatial discretization steps in the simulations and test numerical accuracy and convergence. The time step is $\Delta t \sim 100$ ns. The longitudinal step is $\Delta z \sim 1$ μm for solving Eq. (1). The spatial grids are $\Delta x \sim 0.1\,\mu m$ and $\Delta z \sim 100\,\mu m$ for solving Eq. (2).

### Frequency-domain semi-analytic theory

Equation (5) is a direct generalization of previously studied cases of noise growth upon FM-dominant excitation[17,19]. It ignores any non-phase-matched terms and the self coupling, which occurs on a frequency scale much lower than standard TMI. $T_l$ is the $l$-th eigenmode of the transverse Laplacian $\nabla_\perp^2$ with Dirichlet boundary conditions at the outer cladding surface.

The coefficient in Eq. (6) is $\alpha = 2\eta D(\lambda_s - \lambda_p)/n_c \kappa \lambda_s \lambda_p$, where $\eta \approx 2n_c(dn/dT)$ is the thermo-optical constant, and $D = \kappa/\rho C$ is the heat diffusion constant.

To extract the effective thermo-optical susceptibiltiy from the time-domain simulation of five-mode excitation ($M_e = 5$) in Fig. 3a, we perturb only one mode at a time and examine the output-power fluctuation of that mode. From the time-domain simulation data, we first calculate the noise power in each mode at position $z$, $\tilde{P}_m(z)$, by taking the ratio between the temporal variance and mean of the mode power. We then calculate the relative intensity noise (RIN) of mode $m$ at $z$ as $\text{RIN}_m(z) = \tilde{P}_m(z)/P_m(z)$. The effective thermo-optical susceptibility $\bar{\chi}$ is calculated by taking the log of noise amplification factor (given as the ratio of the output and input RINs), dividing by the total extracted power, and averaging over the five modes, $\bar{\chi} = \langle \bar{\chi}_m \rangle_m = \langle \ln[\text{RIN}_m(L)/\text{RIN}_m(0)] \rangle_m / \sum_m [P_m(L) - P_m(0)]$. This formula is obtained by inverting Eq. (5). For comparison, we calculate the effective thermo-optical susceptibility under the FM-dominant excitation from the fluctuation of the HOM in a two-mode waveguide, $\bar{\chi}(M_e \approx 1) = \chi_{21} = \ln[\text{RIN}_2(L)/\text{RIN}_2(0)]/[P_1(L) - P_1(0)]$. For the theoretical calculation of $\bar{\chi}$, we first calculate pairwise thermo-optical susceptibility $\chi_{mj}(\Omega)$ using Eq. (6), then the mean for mode $m$, $\bar{\chi}_m =$

$\sum_{j \neq m} \chi_{mj}/M_e$, finally average $\overline{\chi}_m$ over $m$. Since only the noise power at downshifted frequency ($\omega_0 - \Omega$) experiences growth due to the thermo-optical effect, we plot only the downshifted spectrum of both theoretical and numerical $\overline{\chi}$ in Fig. 3a, and the perturbation frequency $\Omega$ corresponds to the frequency downshift.

## Acknowledgements

We thank Stephen Warrensmith, Ori Henderson-Sapir, Linh Viet Nguyen, Heike Ebendorff-Heidepriem, and David Ottaway at The University of Adelaide and Peyman Ahmadi at Coherent for stimulating discussions. We acknowledge the computational resources provided by the Yale High Performance Computing Cluster (Yale HPC). The work is supported by the Air Force Office of Scientific Research (AFOSR) under Grant No. FA9550-20-1-0129.

## Author contributions

H.C. conceived the idea and initiated the research. C.-W.C. developed numerical codes with assistance from Y.E., conducted time-domain simulations, and analyzed data, under supervision of H.C. and in collaboration with K.W. K.W. developed frequency-domain theory and performed theoretical analysis, under supervision of A.D.S and in collaboration with C.-W.C. All authors contributed to manuscript writing and editing .

## Conflict of interest

The authors declare no competing interests.